\documentclass[aps,prl,superscriptaddress,twocolumn]{revtex4-1}
\usepackage{amsmath}    
\usepackage{graphicx}   
\usepackage{array}
\usepackage[dvipdfx,cmyk]{xcolor}     
\usepackage[caption=false]{subfig}  
\usepackage{dsfont,bm}

\usepackage[breaklinks=true]{hyperref}

\hypersetup{
  colorlinks   = true, 
  urlcolor     = blue, 
  linkcolor    = blue, 
  citecolor   = green, 
  pdfborder = {0 0 0}
}

\newcommand{\be}{\begin{equation}}
\newcommand{\ee}{\end{equation}}
\newcommand{\ben}{\begin{eqnarray}}
\newcommand{\een}{\end{eqnarray}}
\newcommand{\bes}{\begin{subequations}}
\newcommand{\ees}{\end{subequations}}
\newcommand{\bF}{\begin{figure}}
\newcommand{\eF}{\end{figure}}

\newcommand{\ket}[1]{\vert{#1}\rangle}

\begin{document}
\title{Precision metrology using weak measurements}

\author{Lijian Zhang}
\email{lijian.zhang@nju.edu.cn}
\affiliation{National Laboratory of Solid State Microstructures and College of Engineering and Applied Sciences, Nanjing University, Nanjing 210093, China}
\affiliation{Collaborative Innovation Center of Advanced Microstructures, Nanjing University, Nanjing 210093, China.}
\affiliation{Max Planck Institute for Structure and Dynamics of Material, Hamburg 22761, Germany}

\author{Animesh Datta}
\affiliation{Department of Physics, University of Warwick, Coventry CV4 7AL, United Kingdom}
\affiliation{Clarendon Laboratory, Department of Physics, University of Oxford, Oxford OX1 3PU, United Kingdom}

\author{Ian A. Walmsley}
\affiliation{Clarendon Laboratory, Department of Physics, University of Oxford, Oxford OX1 3PU, United Kingdom}

\date{\today}
\begin{abstract}

Weak values and measurements have been proposed as means to achieve dramatic enhancements in metrology based on the greatly increased range of possible measurement outcomes. Unfortunately, the very large values of measurement outcomes occur with highly suppressed probabilities. This raises three vital questions in weak-measurement-based metrology, namely, (Q1) Does post-selection enhance the measurement precision~? (Q2) Does weak measurement offer better precision than strong measurement~? (Q3) Is it possible to beat the standard quantum limit or to achieve the Heisenberg limit with weak measurement using only classical resources~? We analyse these questions for two prototypical, and generic, measurement protocols and show that while the answers to the first two questions are negative for both protocols, the answer to the last is affirmative for measurements with phase-space interactions, and negative for configuration space interactions. Our results, particularly the ability of weak measurements to perform at par with strong measurements in some cases, are instructive for the design of weak-measurement-based protocols for quantum metrology.


\end{abstract}


\maketitle


Weak measurements reveal partial information about a quantum state without ``collapsing'' it. This is done by coupling a measurement apparatus (MA) feebly to a test quantum system (QS), the dynamics of which is of interest. A procedure involves probing the QS at an intermediate stage between a pre-selected prepared state and a post-selected state which typically has little overlap with the prepared state~\cite{PhysRevLett.60.1351}. A subsequent projective measurement on the MA yields an outcome known as the ``weak value". The fact that the weak value may lie outside the spectrum of the measurement operator leads itself to some interesting results. This phenomena has been used to study numerous quantum effects~\cite{PhysRevLett.74.2405, Yokota2009, lundeen:020404, ruskov:200404, williams:026804, goggin_09, palacios2010experimental, PhysRevLett.92.043601, PhysRevLett.93.203902, PhysRevLett.109.100404, chen2013experimental, Molmer2001151, Resch2004, Mir2007, kocsis2011observing} as well as to reconstruct the wavefunctions of quantum states~\cite{lundeen2011direct, PhysRevLett.108.070402, Salvail2013, wu2013state}.



Weak values may dramatically amplify the small perturbations of the meter state arising from the coupling between the QS and MA~\cite{PhysRevLett.66.1107, susa2012optimal, RevModPhys.86.307}. This amplification makes weak measurements potentially useful in estimating the coupling strength with enhanced precision~\cite{Hosten2008, dixon:173601, brunner_2009, 2013arXiv1306.4768X, PhysRevLett.107.133603, PhysRevLett.106.080405, PhysRevLett.109.013901}. Yet, the amplification effect of weak measurement comes at the cost of a reduced rate at which data can be acquired due to the requirement to select almost orthogonal pre- and post-selected states of the QS. This leads to a majority of trials being ``lost". Thus, the central question is whether the amplification effect of a weak measurement can overcome the corresponding reduction in the occurrence of such events to provide an estimation at a precision surpassing conventional techniques. This issue has garnered substantial interest recently~\cite{starling:041803, PhysRevA.84.052111}, in particular the amplification of information~\cite{2013arXiv1306.2409T, 2013arXiv1307.4016F, combes2013probabilistic} and their role in alleviating technical imperfections~\cite{PhysRevLett.107.133603, PhysRevA.87.012115, PhysRevX.4.011032, PhysRevX.4.011031, 2014arXiv1402.0199V}. However, an unequivocal agreement as to the ultimate efficacy of weak measurements in precision metrology is still lacking. Our endeavour in this work is to provide such an answer in the ideal scenario (i.e. without technical imperfections).



In this Letter, we show that post-selection does not enhance the precision of estimation, that weak measurements do not offer better precision relative to strong measurements, and that it is possible to beat the standard quantum limit and to achieve Heisenberg limit of quantum metrology with weak measurements using only classical resources. These apparently contradictory conclusions arise from a complete consideration of where the maximum information resides in the weak measurement protocol. Our results are valid both for single-particle MA states, in which the QS couples to a continuous degree of freedom of the MA, and for multi-particle states of a bosonic MA. Although in both cases the MA may have similar mathematical representations, the degrees of freedom involved are different, and therefore the scaling of the precision is different and in consequence analyzed separately. Our analysis properly counts the resources involved in the measurement process, enabling us to compare the precision of different measurement strategies and strengths using tools of classical and quantum Fisher information. Weak measurements have a rich structure, and offer some prospects for novel strategies for quantum-enhanced metrology. Nonetheless, we show that a new approach is required to harness this potential.


\textit{Framework :} Our aim is to estimate a parameter associated with the interaction between two systems. We focus on the situation that one of them, the QS, is a two-state system with eigenstates ${|-1\rangle, |+1\rangle}$ of an observable $\hat{S}$ with corresponding eigenvalues -1 and 1. The initial (pre-selected) state of the QS is prepared as $|\psi_i\rangle = \cos (\theta_i/2)|-1\rangle + \sin (\theta_i/2) e^{i\phi_i} |+1\rangle$. The initial state of the other system, the MA, is $\ket{\Phi_i}$. The coupling strength $g$ which is to be estimated appears in the Hamiltonian $H = -g \delta(t-t_0) \hat{S} \hat{M}$ coupling MA to QS, where $\hat{M}$ is an observable of the MA. After this interaction, the joint state of the MA and the QS is
\begin{equation}
|\Psi_j\rangle = \cos \frac{\theta_i}{2}|-1\rangle |\Phi_{-g}\rangle + \sin \frac{\theta_i}{2} e^{i\phi_i} |+1\rangle |\Phi_{+g}\rangle,
\label{eq:state_joint}
\end{equation}
where $|\Phi_{\pm g}\rangle = \exp(\mp i g \hat{M}) \ket{\Phi_i}$. Post-selecting the QS in state $|\psi_f\rangle = \cos (\theta_f/2)|-1\rangle + \sin (\theta_f/2) e^{i\phi_f} |+1\rangle$ leads to the MA state $|\Phi_d\rangle = \left(\gamma_{d}^{-}|\Phi_{-g}\rangle + \gamma_{d}^{+}|\Phi_{+g}\rangle \right)/\sqrt{p_d}, $ with $\gamma_{d}^{-} = \cos (\theta_i/2) \cos (\theta_f/2)$, $\gamma_{d}^{+} = \sin (\theta_i/2) \sin (\theta_f/2) \exp(i\phi_0)$ and $\phi_0 = \phi_i - \phi_f$. The probability of successful post-selection, i.e., of obtaining $\ket{\Phi_d}$ is $p_d$. When the post-selection fails (with probability $p_r = 1-p_d$), the MA state, which is not considered in the original protocol and is often ignored in experiments, is
$ |\Phi_r\rangle = \left(\gamma_{r}^{-}|\Phi_{-g}\rangle + \gamma_{r}^{+} |\Phi_{+g}\rangle \right)/\sqrt{1-p_d},$ where $\gamma_{r}^{-} = \cos (\theta_i/2) \sin (\theta_f/2)$, $\gamma_{r}^{+} = - \sin (\theta_i/2) \cos (\theta_f/2) \exp(i\phi_0)$. Repeating the pre-selection-coupling-post-selection process $N$ times, yields $Np_d$ copies of $\ket{\Phi_d}$ and $N(1-p_d)$ copies of $\ket{\Phi_r}$. The best attainable precision in estimating $g$ is given by the Cram\'{e}r-Rao bound $\Delta^2 g \geq 1/(NF_{tot})$~\cite{PhysRevLett.72.3439}, where $F_{tot}$ is the sum total of the classical and quantum Fisher information (FI) contained at different stages of the pre-selection--coupling--post-selection process. Note that the single-parameter Cram\'{e}r-Rao bound, both quantum and classical, can always be attained asymptotically for large $N$ with maximum-likelihood estimation.

Depending on the estimation protocol, $F_{tot}$ may have different values. To date almost all applications of the weak measurement to precision metrology focus on the amplification effect of weak values, which corresponds to considering the information about $g$ contained in $|\Phi_d\rangle$. In this situation $F_{tot} = p_d Q_{d}$, where $Q_{d}$ is the quantum FI (QFI) of $\ket{\Phi_d}$, \textit{i.e.} the maximum FI that can be achieved with the optimal measurement on $\ket{\Phi_d}$, which is a set of projection operators onto the eigenstates of the symmetric logarithmic derivative of $\ket{\Phi_d}$~\cite{PhysRevLett.72.3439}.
$p_d Q_{d}$ can be viewed as the total information in the post-selected meter state. In addition, one may also monitor the failure mode $|\Phi_r\rangle$ to achieve better precision in parameter estimation~\cite{PhysRevA.86.040102, PhysRevLett.110.083605} and state tomography~\cite{wu2013state}. The maximum information in the failure mode is $(1-p_d)Q_r$ where $Q_r$ is the QFI of $\ket{\Phi_r}$. Finally, the distribution $\{p_d,1-p_d\}$ of the post-selection process on QS also contains information about $g$. This distribution yields a classical FI $F_p$ which we refer to as the information in the post-selection process. If we account for all these contributions, we have (see supplementary material for a proof)
\begin{equation}
F_{tot} = p_d Q_d + (1-p_d)Q_r + F_p.
\label{eq:ftot}
\end{equation}
The whole process (post-selection plus measurements on the MA state) is a special case of the global measurement on the joint state $|\Psi_j\rangle$~\cite{supp}, therefore $F_{tot}$ is no larger than the QFI $Q_j$ of $|\Psi_j\rangle$, \textit{i.e.} post-selection cannot increase the precision in estimating $g$. This seemingly straightforward result provides  important insight about the relation between the amplification effect and measurement precision, and allows us to access the rich structures of weak measurement and evaluate their quantum advantages. In particular, we note that $Q_d$ or $Q_r$ alone may be larger than $Q_j$ due to the amplification effect of weak values. Nevertheless this apparent gain of information is completely canceled by the small probability of successful post-selection. Moreover, the post-selection process may contain important information $F_p \geq 0$. This analysis goes beyond previous studies~\cite{2013arXiv1306.2409T} by considering all the contributions to the total information, and thus provides a complete answer to Q1 posed in the abstract. We note that a similar conclusion was independently and contemporaneously reached in~\cite{combes2013probabilistic}. In following sections, we provide answers to Q2 and Q3 in both configuration and phase space interactions.


\textit{Configuration space interactions :} We begin with the most widely used scenario in weak measurement~\cite{PhysRevLett.60.1351, PhysRevLett.66.1107, PhysRevLett.92.043601, PhysRevLett.93.203902, PhysRevLett.66.1107, Hosten2008, dixon:173601, brunner_2009, 2013arXiv1306.4768X, PhysRevLett.106.080405, PhysRevLett.109.013901} where both the QS and MA are single-particle states, possibly in different degrees of freedom of the same particle~\footnote{To be precise, the QS and MA in these implementations are (multi-mode) coherent states. Yet as we will show, the following analysis can be applied with little modification}. In this situation, the MA is normally prepared in a Gaussian superposition state of two conjugate variables
\ben
|\Phi\rangle &=& \int dq \frac{1}{(2\pi\sigma^2)^{1/4}} \exp(-\frac{q^2}{4\sigma^2}) |q\rangle \nonumber \\
            &=& \int dp \frac{(2 \sigma^2)^{1/4}}{\pi^{1/4}} \exp(-\sigma^2 p^2) |p\rangle, \label{eq:meter_gauss}
\een
where $p$ and $q$ are, e.g., momentum and position or time and frequency. The two representations are related via a Fourier transform. The interaction Hamiltonian between the QS and MA is chosen as $H = -g \delta(t-t_0) \hat{S} \hat{q}$. Note that this interaction Hamiltonian entangles the QS with an \textit{external} degree of freedom of the MA. It does not change the particle number distribution in the state of the MA. After the interaction and post-selection, the MA state becomes $|\Phi_k\rangle = \int dp \phi_k(g, p) |p\rangle$ ($k = d,r$) with
\be
\phi_k(g,p) = \frac{(2 \sigma^2)^{1/4}}{\pi^{1/4}\sqrt{p_k}} \left[\gamma_{k}^{-}e^{-\sigma^2 (p+g)^2} + \gamma_{k}^{+} e^{-\sigma^2 (p-g)^2}\right]. \label{eq:phi_phys}
\ee
The probability of successful post-selection is
\be
p_d = \frac{1+\cos\theta_i \cos\theta_f + \sin\theta_i \sin\theta_f \cos\phi_0 e^{-2s^2}}{2},
\label{eq:ps_phys}
\ee
with $s = g\sigma$ characterising the measurement strength. With Eqns. (\ref{eq:phi_phys}, \ref{eq:ps_phys}) we can estimate $Q_d$, $Q_r$ and $F_p$~\cite{supp}.
\ben
Q_d & = & \frac{4\sigma^2}{p_d} \left[p_d + S\left(2s^2-1\right)-\frac{1}{p_d}S^2 s^2\right], \nonumber \label{eq:qm_phys} \\
Q_r & = & \frac{4\sigma^2}{1-p_d} \left[1-p_d - S\left(2s^2-1\right)-\frac{1}{1-p_d}S^2 s^2\right], \nonumber \label{eq:qf_phys} \\
F_p & = & \frac{4\sigma^2s^2 S^2}{p_d(1-p_d)}, \label{eq:fp_phys}
\een
where $S=e^{-2s^2}\sin\theta_i \sin\theta_f \cos\phi_0$. Further the QFI of the joint meter-system state before post-selection is $Q_j = 4\sigma^2$. We can now calculate $F_{tot}$ for different estimation strategies. In particular, if we take into account of all the contributions in Eq.~(\ref{eq:ftot}), we have $F_{tot} = Q_j$, \textit{i.e.} we achieve the maximal precision. A commonly employed strategy retains only the information in the successfully post-selected meter state. In this case, the complicated functional form of $F_{tot}=p_d Q_d$ demands numerical maximization over $\psi_i$ and $\psi_f$. Nonetheless, some limits that may be obtained analytically allow us to answer Q2. In the weak measurement limit, defined as $s\rightarrow 0$
\begin{equation}
p_d Q_d = 2 \sigma^2 (1+\cos\theta_i \cos\theta_f - \sin\theta_i \sin\theta_f \cos\phi_0),
\label{eq:fi_MA_weak}
\end{equation}
the maximum value of which is $4\sigma^2$, attained when either $\theta_i = -\theta_f$ and $\phi_0 =0$ or $\theta_i = \theta_f$ and $\phi_0 = \pi$. Interestingly, this does not coincide in general with the situation when the weak value is the largest which requires $p_d = |\langle \psi_i | \psi_f \rangle|^2 \rightarrow 0$~\cite{PhysRevA.84.052111}. In the limit of strong measurement when $s \gg 1$,
\begin{equation}
p_d Q_d = 2 \sigma^2 (1+\cos\theta_i \cos\theta_f),
\label{eq:fi_MA_strong}
\end{equation}
which also attains the maximum of $4\sigma^2,$ but for the situation that both pre- and post-selected states are $\ket{+1}$ or $\ket{-1}$. In both these limits, $p_d Q_d = Q_j$, $F_p=0$ and $Q_r = 0$.

More generally, non-Gaussian MA states also achieve this precision (see the supplementary material for proof). This may be relevant to recent experiments that exploit this resource~\cite{shomroni2013demo, PhysRevLett.109.040401}. The conclusion is that when the uncertainty of the meter state $\sigma$ is fixed, the precision in the weak measurement limit, that is, to estimate a small parameter $g$ through pre-selection-coupling-post-selection, is no better than that in the strong measurement limit, that is, when the coupling parameter is large. However, if the parameter to be estimated is fixed, the precision is always better if we use a meter state with larger $\sigma$, as is evident in Eqns.~(\ref{eq:fi_MA_weak}, \ref{eq:fi_MA_strong}) and $F_{tot}$ since the FIs are proportional to $\sigma^2$. This answers Q2 for the configuration-space-interaction scenario. 

This analysis focuses on the effect of the uncertainty in the external degrees of freedom of the MA as in the previous works~\cite{PhysRevLett.66.1107, Hosten2008, dixon:173601, brunner_2009, 2013arXiv1306.4768X, PhysRevLett.106.080405, PhysRevLett.109.013901, starling:041803, PhysRevA.84.052111, PhysRevA.86.040102}, showing that weak measurements may or may not offer an overhead advantage. In quantum metrology, the relevant measure of the resource required to effect a measurement is the average number of photons ($n$) in the MA state. The scaling of the precision of estimation with respect to $n$ is the signature of whether the system is capable of operating beyond the standard quantum limit (in which the FI scales linearly in $n$) and offering genuine quantum advantages. Since the interaction Hamiltonian does not change particle-number distributions, for QS and MA prepared in (multi-mode) coherent states with amplitude $\alpha$, post-selected meter states are also multi-mode coherent states, and the FIs in Eqns. (\ref{eq:fi_MA_weak}, \ref{eq:fi_MA_strong}) pick up an additional factor of $n=|\alpha|^2$. Thus the scalings are at the standard quantum limit. 
This is the answer to Q3 for the configuration-space-interaction scenario.




\textit{Phase-space interactions:} We now consider a scenario that can change the particle-number distribution. 
The initial state of the QS $\ket{\psi_i}$ is the same as before, while the MA is prepared in a coherent state $\ket{\alpha}$. A state-dependent interaction with $\hat{M} = \hat{n}$, where $\hat{n}$ is the particle number operator, leads to~\footnote{See, for instance, Eq. (2) in Ref.~\cite{PhysRevLett.107.133603}, where $g = \phi_0/2$.}
\begin{equation}
|\Psi\rangle = \cos \frac{\theta_i}{2}|-1\rangle |\alpha \rangle + \sin \frac{\theta_i}{2} e^{i\phi_i} |+1\rangle |\alpha e^{i2g}\rangle.
\label{eq:psi_full_2}
\end{equation}
or~\cite{2014arXiv1409.3488J}
\begin{equation}
|\Psi\rangle = \cos \frac{\theta_i}{2}|-1\rangle |\alpha e^{-ig}\rangle + \sin \frac{\theta_i}{2} e^{i\phi_i} |+1\rangle |\alpha e^{ig}\rangle.
\label{eq:psi_full}
\end{equation}
Both states have the same precision in estimating $g$ when $n$ is large. In the following, we focus on the symmetric form in Eq.~(\ref{eq:psi_full}). The meter states after post-selection are ($k=d,r$) $|\Phi_k\rangle = \left(\gamma_{k}^{-}|\alpha e^{-ig}\rangle + \gamma_{k}^{+} |\alpha e^{ig}\rangle\right)/\sqrt{p_k}. $
The probability of obtaining this state and the FIs are all given in~\cite{supp}. Again, the QFIs are attainable with the optimal measurement on $|\Phi_k\rangle$.


The QFI of the system-meter state in Eq.~(\ref{eq:psi_full}) is~\cite{supp} $ Q_{j} = 4n^2\sin^2\theta_i+4n, $ where $n=|\alpha|^2$ is again the mean photon number (or energy) of the meter state (Similarly, the QFI of the state in Eq.~(\ref{eq:psi_full_2}) is $Q_{j} = 4n^2\sin^2\theta_i + 4n[4\sin^2(\theta_i/2)]$). 
$Q_{j}$ is the maximum amount of information, and can exhibit quantum scaling ($\sim n^2$) depending on the initial system state. The expression for $Q_{j}$ immediately suggests that $\theta_i=0,\pi$ will never provide a better-than-classical scaling. These are the two cases when the initial state is an eigenstate of $\hat{S}$, so that no entanglement is generated between the QS and MA. Indeed, for $|\psi_i\rangle = \ket{\pm 1},$ $p_d Q_d  = 2n(1\pm \cos\theta_f),$ $(1-p_d)Q_{r}=2n(1\mp \cos\theta_f)$ and $F_p=0.$ Thus, $F_{tot}=4n,$ but the information may be equally shared between the successful and the failed post-selection mode. This is important since the failed post-selection mode is generally discarded completely~\cite{PhysRevLett.66.1107, Hosten2008, dixon:173601, brunner_2009, 2013arXiv1306.4768X, PhysRevLett.106.080405, PhysRevLett.109.013901, PhysRevLett.107.133603}.

In contrast, the maximal $Q_j$ is found for $\theta_i=\pi/2$. We immediately find that $\theta_f=0,\pi$ provides no better than classical scalings either. Thus, we set $\theta_f=\pi/2$ as well, and find that as $g\rightarrow 0,$ it leads to
\be
F_p=4n^2.
\label{eq:heisenberg}
\ee
This result shows that quantum-enhanced scaling can be attained in the sensing of the coupling parameter $g$ in a weak measurement setup. On the other hand, in this same situation the QFIs for both the successful and failed post-selection mode scale classically; $p_d Q_d=4n\sin^2(\phi_0/2)$ and $(1-p_d)Q_r=4n\cos^2(\phi_0/2)$, where $\phi_0 = \phi_i-\phi_f$. This shows that $p_d Q_d$ achieves its maximum when $\phi_0 \rightarrow \pi$, \textit{i.e.} $\psi_i$ and $\psi_f$ are orthogonal. Note also that if we take into account of all the contributions we have $F_{tot}=Q_{j}.$ This is a particularly interesting situation since most, if not all, earlier experiment considered only the information $Q_d$ contained in the successfully post-selected MA state. Yet, as our calculation shows, the post-selection process has much more information, and indeed scales at the Heisenberg limit. The parameter $g$ can be estimated with the precision derived in Eq.~(\ref{eq:heisenberg}) from the statistics of the success/failure of the post-selection using a maximum likelihood estimator.

\begin{figure}[h]
\includegraphics[width=8cm]{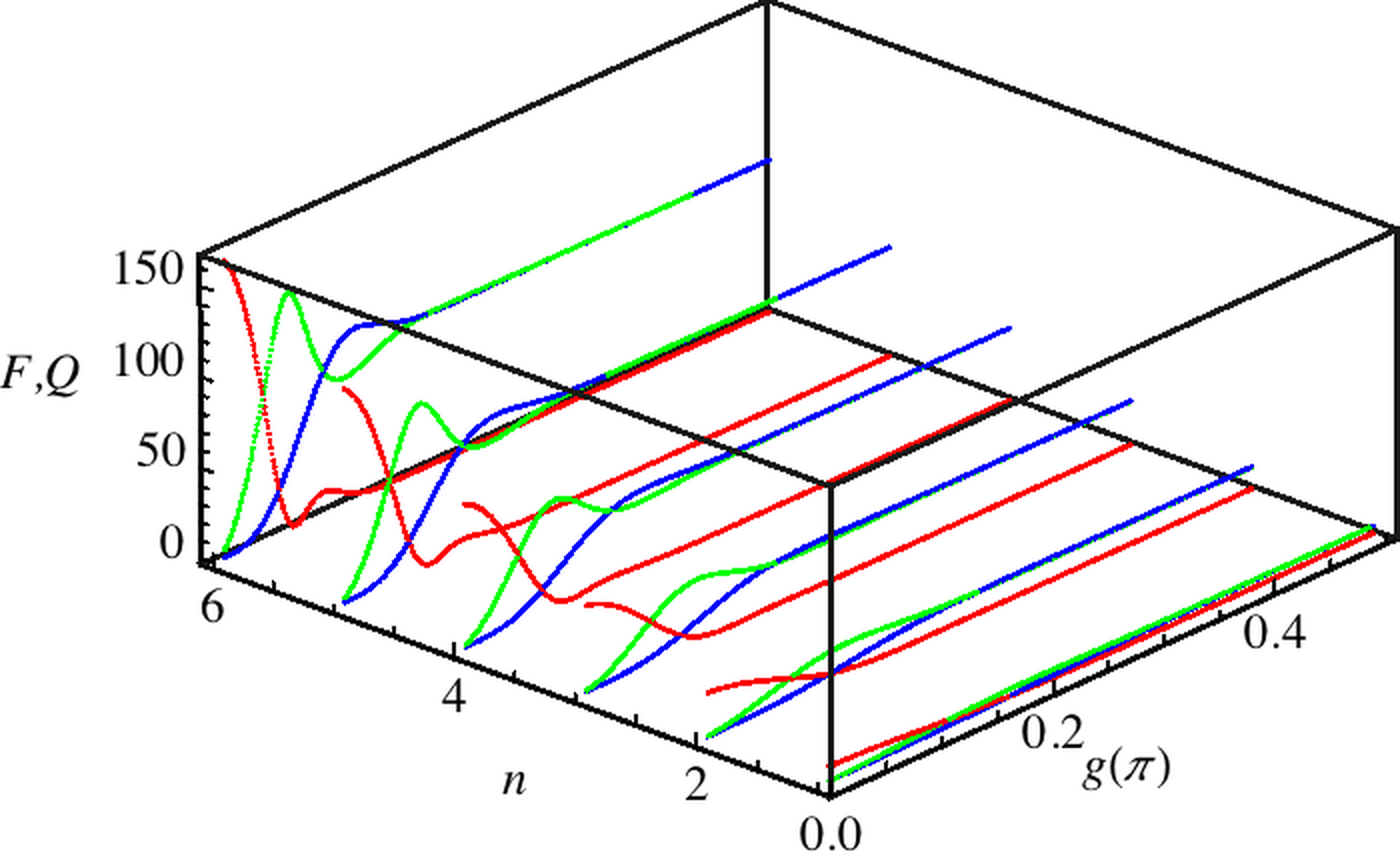}
\caption{Contributions to the total information from the three constituents in the conditional-phase-rotation scenario with pre- and post-selected QS state $\psi_i = (\ket{-1}+\ket{+1})/\sqrt{2}$ and $\psi_f = (\ket{-1}-\ket{+1})/\sqrt{2}$, and initial MA state $|\alpha\rangle$. Red : $F_p,$ Green : $ p_d Q_d,$ Blue : $(1-p_d)Q_r.$ The sum of three quantities $F_{tot}$ equals $Q_j,$ the total QFI of the joint system-meter state which is $4n^2+4n$ and $n = |\alpha|^2$.}
\label{fig:allqfi}
\end{figure}

\begin{figure}[h]
\includegraphics[width=8cm]{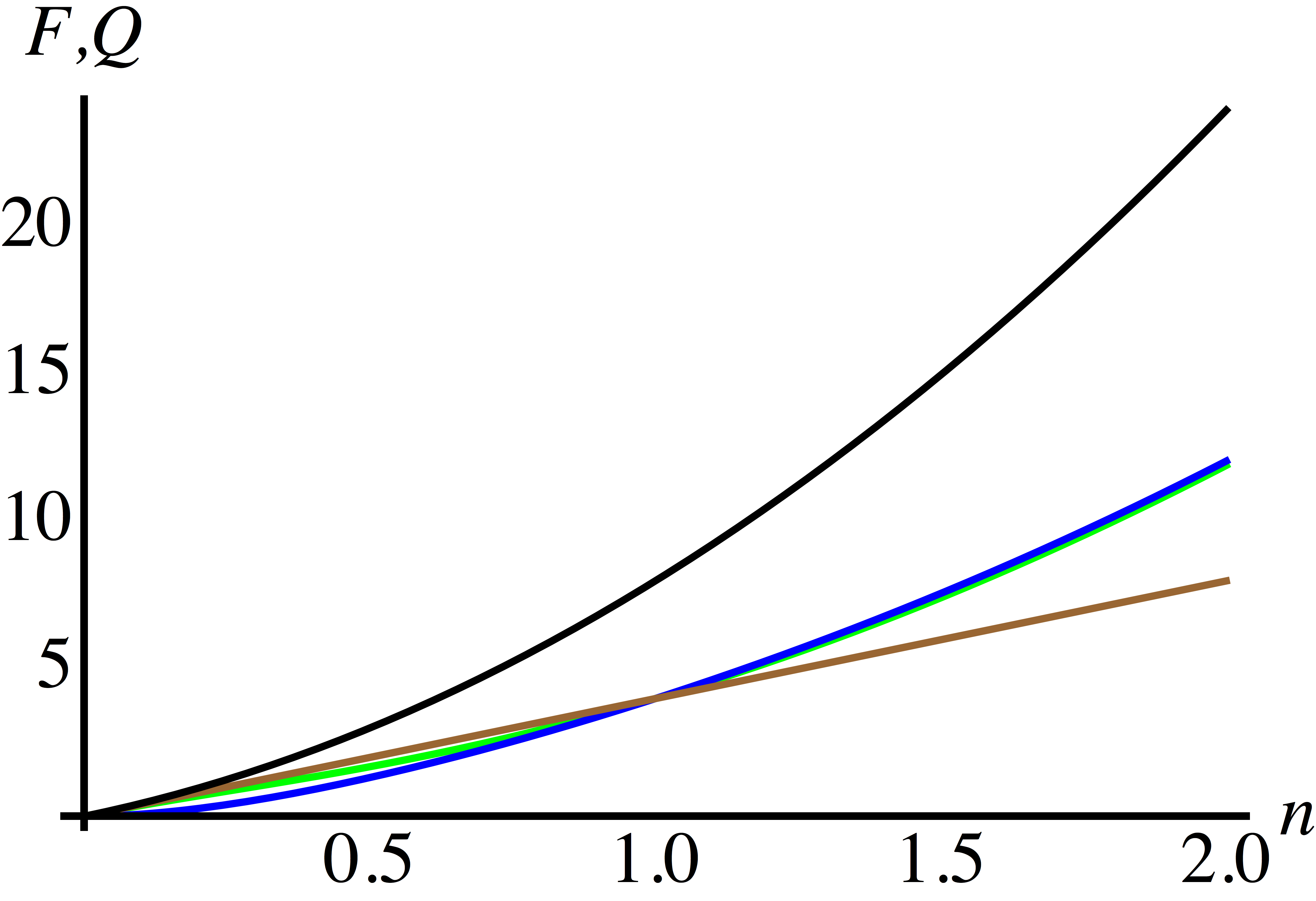}
\caption{Classical and quantum FIs for $g=\pi/2.$ in the conditional phase rotation scenario with with $\psi_i = (\ket{-1}+\ket{+1})/\sqrt{2}$ and $\psi_f = (\ket{-1}-\ket{+1})/\sqrt{2}$, and initial MA state $|\alpha\rangle$. Green : $ p_d Q_d,$ Blue : $(1-p_d)Q_r,$ Black: $Q_{j}=4n^2+4n,$ and Brown: Classical scaling of $4n$. $F_{p}$ is not shown since it is 0. The green and blue lines add up to the black line.}
\label{fig:qfi}
\end{figure}

For interaction strengths $g > 0,$ the contributions of the different terms in $F_{tot}$ change. In Fig.~(\ref{fig:allqfi}), we plot the FI and QFIs contributing to $F_{tot}$ for $\phi_0=\pi.$ Exploiting a symmetry of our model, we only plot the results in $g=\{0,\pi/2\}$. As shown earlier, for $g \rightarrow 0$ the main contribution comes from $F_p,$ the classical FI in the post-selection distribution. As $g$ increases, $F_p$ falls, and the information in the post-selected states for both successful and failed QS measurement outcomes rises. For $g=\pi/2,$ we plot the contributions in greater detail in Fig.~(\ref{fig:qfi}) for $\phi_0=\pi.$ For this case, $F_p=0$ while $ (1-p_d)Q_r$, $p_d Q_d$ are almost equal. Indeed the difference in the QFIs decreases with $n$, as
$ p_d Q_d - (1-p_d)Q_r = -4n(n-1) \exp(-2n).$ For $n \gg 1,$ up to a small exponential correction, there is thus as much information in the successful post-selection mode as in the failed mode, and both of them scale better than the classical scaling. 
In all cases, the total $F_{tot}$ still matches the maximum QFI attainable, that is $Q_{j}.$
These results provide answers to Q2 and Q3 for the conditional-phase-shift scenario.

\textit{Discussion and Conclusions :} It is perhaps unsurprising that the Heisenberg limit for estimating the coupling parameter $g$ in the conditional-phase-shift interaction can be attained when the system-meter coupling is strong, since in that case, the post-selected MA states are Schr\"odinger-cat states. That is, the measurement protocol produces highly non-classical states in the joint system. In the case of weak coupling ($g\rightarrow 0$), however, the the post-selected MA states are classical, and the Heisenberg scaling arises only in the post-selection process itself. How this conditioning step using a classical MA state achieves a precision beyond the standard quantum limit is therefore an interesting open question.



Our calculations show that not only the failed post-selection mode but the post-selection process itself contains useful information. The analysis provide answers to three long-standing questions in the study of weak measurement posed in the abstract: (A1) Post-selection can not enhance the measurement precision even when all the contributions are taken into account; (A2) For equal resources, weak measurement does not give improved precision over strong measurement, when both measurements are optimized. In particular, this result applies to all previous experiments that have explored weak-measurement enhancements to precision metrology.  (A3) Weak measurement that modifies the particle number distribution of the meter state can yield quantum-enhanced precision though no non-classical states need be involved. These results highlight the rich structure of the weak measurement and shed new light on both the understanding of quantum measurement and the development of new technologies for practical quantum metrology.



\begin{acknowledgments}
We thank M. Barbieri for several useful comments on the manuscript. This work was supported by National Basic Research Programme of China (No. 2011CBA00205), the Engineering and Physical Sciences Research Council (EP/H03031X/1, EP/K034480/1, EP/K04057X/1, EP/M01326X/1 and EP/M013243/1), the Air Force Office of Scientific Research (European Office of Aerospace Research and Development), and the Priority Academic Program Development of Jiangsu Higher Education Institutions. LZ acknowledges support from Alexander von Humboldt Foundation and 1000 Youth Fellowship Program of China.
\end{acknowledgments}

%

\clearpage
\section{Supplementary Material}
\subsection{Derivation of $F_{tot}$}
\label{sec:f_tot}

The quantum FI $Q_k$ of $|\Phi_k\rangle$ ($k=d,r$) is given by
\be
Q_k = 4\left[\left(\frac{d\langle\Phi_k|}{dg}\right) \left(\frac{d|\Phi_k\rangle}{dg}\right) - \left|\langle \Phi_k |\left(\frac{d|\Phi_k\rangle}{dg}\right)\right|^2 \right].
\label{eq:qfi}
\ee
$Q_k$ can be achieved with the optimal POVMs. Assume the optimal measurement for $|\Phi_d\rangle$ is $\{\Pi^{d}_{1} \cdots \Pi^{d}_{V}\}$ with the probabilities of each outcome $\{P(1|\textrm{detect}), \cdots ,P(V|\textrm{detect})$, where
\be
P(v|\textrm{detect}) = \langle \Phi_d |\Pi^{d}_{v} |\Phi_d\rangle, \textrm{ for } v=1\cdots V.
\ee
Then we have
\be
Q_d = \sum_{v=1}^{V} \frac{1}{P(v|\textrm{detect})} \left(\frac{d(P(v|\textrm{detect}))}{dg}\right)^2.
\ee
Similarly, the optimal measurement for $|\Phi_r\rangle$ is $\{\Pi^r_{1} \cdots \Pi^r_{W}\}$ with the probabilities of each outcome $\{P(1|\textrm{reject}), \cdots ,P(W|\textrm{reject})\}$. Then post-selection on the QS state followed by the optimal measurement on the MA states can be considered as a POVM performed on the joint state $\ket{\Psi}$ with $\{|\psi_f\rangle\langle\psi_f|\otimes\Pi^{d}_{1}, \cdots |\psi_f\rangle\langle\psi_f|\otimes\Pi^{d}_{V}, |\psi_f^{\perp}\rangle\langle\psi_f^{\perp}|\otimes\Pi^{r}_{1}, \cdots, |\psi_f^{\perp}\rangle\langle\psi_f^{\perp}|\otimes\Pi^{r}_{W}\}$, where $|\psi_f^{\perp}\rangle$ is the state of QS when post-selection fails. The probabilities associated with each outcome are $\{p_d\times P(1|\textrm{detect}), \cdots ,p_d\times P(V|\textrm{detect}), (1-p_d)\times P(1|\textrm{reject}), \cdots ,(1-p_d)\times P(W|\textrm{reject})\}$. The Fisher information is given by
\begin{eqnarray}
F_{tot} & = & \sum_{v=1}^{V} \frac{1}{p_d P(v|\textrm{detect})} \left(\frac{d(p_d P(v|\textrm{detect}))}{dg}\right)^2 \nonumber \\
 & & + \sum_{w=1}^{W} \frac{1}{(1-p_d) P(w|\textrm{reject})} \left(\frac{d((1-p_d) P(w|\textrm{reject}))}{dg}\right)^2 \nonumber \\
& = & p_d Q_d + (1-p_d) Q_r + F_p,
\label{eq:fisher_total}
\end{eqnarray}
where
\be
F_p = \frac{1}{p_d}\left(\frac{d p_d}{dg}\right)^2+\frac{1}{1-p_d}\left(\frac{d(1-p_d)}{dg}\right)^2.
\ee
If we ignore the meter state when the post-selection fails, the whole process can still be considered as a POVM performed on the joint state with $\{|\psi_f\rangle\langle\psi_f|\otimes\Pi^{d}_{1}, \cdots, |\psi_f\rangle\langle\psi_f|\otimes\Pi^{d}_{V}, |\psi_f^{\perp}\rangle\langle\psi_f^{\perp}|\otimes \hat{I}\}$. The probabilities associated with each outcome are $\{p_d P(1|\textrm{detect}), \cdots ,p_d P(V|\textrm{detect}), 1-p_d\}$. The total FI is given by
\be
F_{tot} = p_d Q_d + F_p.
\ee

\subsection{Configuration space interactions with arbitrary MA state}

We generalize the situation considered in the manuscript to arbitary MA states
\be
|\Phi\rangle = \int dp f(p) |p\rangle,
\ee
where the probability amplitude $f(p)$ satifies the conditions 
\ben
& & \int_{-\infty}^{\infty} |f(p)|^2 dp = 1, \label{eq:normlize} \\
& & |f(p)| \rightarrow 0 \textrm{ when } p \rightarrow \pm\infty, \label{eq:f_infty} \\
& & |f'(p)| \rightarrow 0 \textrm{ when } p \rightarrow \pm\infty. \label{eq:f_derive_infty}
\een
After the interaction between the QS and MA, the joint state is
\ben
|\Psi_j\rangle & = & \cos\frac{\theta_i}{2} |-1\rangle \int dp f(p+g) |p\rangle \nonumber \\
&& + \sin\frac{\theta_i}{2} e^{i\phi_i} |+1\rangle \int dp f(p-g) |p\rangle.
\label{eq:joint_state_arbitary}
\een
Using the conditions that 
\ben
\frac{\partial f(p+g)}{\partial g} & = & f'(p+g), \label{eq:deri_pos_g} \\
\frac{\partial f(p-g)}{\partial g} & = & -f'(p-g), \label{eq:deri_neg_g}
\een
we have
\ben
\frac{d|\Psi_j\rangle}{dg} & = & \cos\frac{\theta_i}{2} |-1\rangle \int dp f'(p+g) |p\rangle \nonumber \\
&& - \sin\frac{\theta_i}{2} e^{i\phi_i} |+1\rangle \int dp f'(p-g) |p\rangle,
\een
then 
\ben
\left(\frac{d\langle\Psi_j|}{dg}\right) \left(\frac{d|\Psi_j\rangle}{dg}\right) & = & \cos^2 \frac{\theta_i}{2} \int dp |f'(p+g)|^2  \nonumber \\ 
& & + \sin^2 \frac{\theta_i}{2} \int dp |f'(p-g)|^2 \nonumber \\
& = & \int dp |f'(p)|^2, 
\een
where we have used $\int_{-\infty}^{\infty} dp |f'(p+g)|^2 = \int_{-\infty}^{\infty} dp |f'(p-g)|^2 = \int_{-\infty}^{\infty} dp |f'(p)|^2$. Similarly we have
\ben
\langle \Psi_j| \left(\frac{d|\Psi_j\rangle}{dg}\right) &=& \cos^2 \frac{\theta_i}{2} \int dp \tilde{f}(p+g) f'(p+g) \nonumber \\
& & - \sin^2 \frac{\theta_i}{2} \int dp \tilde{f}(p-g) f'(p-g) \nonumber \\
& = & \cos \theta_i \int dp \tilde{f}(p) f'(p), 
\een
where $\tilde{f}(p)$ is the conjugate of $f(p)$. Again we have used $\int_{-\infty}^{\infty} dp \tilde{f}(p+g) f'(p+g) = \int_{-\infty}^{\infty} dp \tilde{f}(p-g) f'(p-g) = \int_{-\infty}^{\infty} dp \tilde{f}(p) f'(p)$.
So we have the quantum FI of the joint meter-system state
\ben
Q_j &=& 4\left[\left(\frac{d\langle\Psi_j|}{dg}\right) \left(\frac{d|\Psi_j\rangle}{dg}\right) - \left|\langle \Psi_j |\left(\frac{d|\Psi_j\rangle}{dg}\right)\right|^2 \right] \nonumber \\
&=& 4\left( \int dp |f'(p)|^2 - \cos^2 \theta_i \left| \int dp \tilde{f}(p) f'(p) \right|^2\right).
\label{eq:qj_arbi_meter}
\een
From Eq.~(\ref{eq:qj_arbi_meter}) we can see that $Q_j$ is independent of the value of $g$, \textit{i.e.} the measurement strength. It is worth to investigate this result a bit further. Since
\begin{widetext}
\ben
\int dp \left[\tilde{f} (p) f'(p) + f(p) \tilde{f}'(p)\right] & = & \int dp \left[\tilde{f} (p+g) f'(p+ g) + f(p+g) \tilde{f}'(p+g)\right] \nonumber \\
&=& \int dp \left[\tilde{f} (p+g) \frac{\partial f(p+ g)} {\partial g} + f(p+g) \frac{\partial\tilde{f}(p+g)}{\partial g}\right] \nonumber \\
&=& \frac{d \int dp |f(p+g)|^2}{dg}, \nonumber \\
& = & 0
\een
\end{widetext}
we have 
\be
\int dp \tilde{f}(p) f'(p) = - \int dp f(p) \tilde{f}'(p), 
\label{eq:int_f_f_deri}
\ee
\textit{i.e.} $\int dp \tilde{f}(p) f'(p)$ is either 0 or an imaginary number. In particular, if $f(p)$ is a real function, $\int dp \tilde{f}(p) f'(p)=0$, and $Q_j$ is independent of the choice of the initial QS state. If $\int dp \tilde{f}(p) f'(p)$ is not zero, $Q_j$ reaches its maximum when $\cos \theta_i = 0$, \textit{i.e.} with the initial QS state $|\psi_i\rangle = (|-1\rangle \pm e^{i\phi_i} |+1\rangle)/\sqrt{2}$. 

Now we consider the effect of post-selection, after which the MA state becomes $|\Phi_k\rangle = \int dp \phi_k (g,p) |p\rangle$ ($k = d,r$) with 
\be
\phi_k (g,p) = \frac{1}{\sqrt{p_k}} \left[ \gamma_k^{-} f(p+g) + \gamma_k^{+} f(p-g)\right] = \frac{1}{\sqrt{p_k}} \vartheta_k (p,g),
\ee
where we define $\vartheta_k (p,g) = \gamma_k^{-} f(p+g) + \gamma_k^{+} f(p-g) $ for the later analysis. The probability of successful post-selection is
\begin{widetext}
\be
p_d = \int dp |\vartheta_d (p,g)|^2 = \frac{1+\cos\theta_i \cos\theta_f + \sin\theta_i \sin\theta_f \textrm{Re}\left(\int dp \tilde{f}(p-g) f(p+g) e^{i\phi_0} \right)}{2}
\label{eq:success_prob_arti_meter}
\ee
\end{widetext}
and the probability that the post-selection fails is $p_r = |\int dp \vartheta_r (p,g)|^2 = 1-p_d$. We have
\ben 
p_k Q_k & = & 4 \left[ \int dp \left| \frac{\partial \vartheta_k (p,g)}{\partial g} \right|^2 \right. \nonumber \\
& &\left. - \frac{1}{p_k} \left| \int dp \tilde{\vartheta}_k (p,g) \frac{\partial \vartheta_k (p,g)}{\partial g} \right|^2\right], \label{eq:qfi_post_arbi_meter} \\
F_p & = & \sum_{k = d,r} \frac{1}{p_k} \left(\frac{dp_k}{dg} \right)^2. \label{eq:cfi_post_arbi_meter}
\een
Since $p_k = \int dp |\vartheta_k(p,g)|^2$, we have
\be
\frac{dp_k}{dg}  =  \int dp \frac{\partial \tilde{\vartheta}_k(p,g)}{\partial g} \vartheta_k (p,g)  
+ \int dp \tilde{\vartheta}_k (p,g) \frac{\partial \vartheta_k(p,g)}{\partial g}. 
\label{eq:dp_dg}
\ee
\newpage
Substituting Eq. (\ref{eq:dp_dg}) into Eq. (\ref{eq:cfi_post_arbi_meter}), and summing over all the contributions to the total Fisher information, we have
\begin{widetext}
\be
F_{tot} = \sum_{k = d, r} \left\{ 4 \int dp \left| \frac{\partial \vartheta_k (p,g)}{\partial g} \right|^2 + \frac{1}{p_k}\left[ \int dp \left( \frac{\partial \tilde{\vartheta}_k(p,g)}{\partial g} \vartheta_k (p,g) - \tilde{\vartheta}_k (p,g) \frac{\partial \vartheta_k(p,g)}{\partial g}\right) \right]^2 \right\}.
\label{eq:ftot_arbi_meter}
\ee
\end{widetext}
Here we are interested in the situations in the weak and strong measurement limit, \textit{i.e.} $g\to 0$ and $g \to \infty$. The results for a general $g$ will be given elsewhere. 

In the weak measurement limit with $g \to 0$, we have 
\begin{widetext} 
\ben
p_d & = & \frac{1+\cos\theta_i \cos\theta_f + \sin\theta_i \sin\theta_f \cos\phi_0}{2}, \label{eq:success_prob_arbi_meter_weak} \\
p_r & = & \frac{1-\cos\theta_i \cos\theta_f - \sin\theta_i \sin\theta_f \cos\phi_0}{2}, \label{eq:fail_prob_arbi_meter_weak} \\
\int dp \left| \frac{\partial \vartheta_d (p,g)}{\partial g} \right|^2 & = & \frac{1+\cos\theta_i \cos\theta_f - \sin\theta_i \sin\theta_f \cos\phi_0}{2} \int dp |f'(p)|^2, \label{eq:int_partial_d_weak} \\
\int dp \left| \frac{\partial \vartheta_r (p,g)}{\partial g} \right|^2 & = & \frac{1-\cos\theta_i \cos\theta_f + \sin\theta_i \sin\theta_f \cos\phi_0}{2} \int dp |f'(p)|^2, \label{eq:int_partial_r_weak} \\
\int dp \tilde{\vartheta}_d (p,g) \frac{\partial \vartheta_d(p,g)}{\partial g} & = & \frac{\cos\theta_i + \cos\theta_f + i \sin\theta_i\sin\theta_f \sin\phi_0}{2} \int dp \tilde{f} (p) f'(p), 
 \label{eq:int_f_partial_f_d_weak} \\
\int dp \tilde{\vartheta}_r (p,g) \frac{\partial \vartheta_r(p,g)}{\partial g} & = & \frac{\cos\theta_i - \cos\theta_f - i \sin\theta_i\sin\theta_f \sin\phi_0}{2} \int dp \tilde{f} (p) f'(p),  \label{eq:int_f_partial_f_r_weak} 
\een
\end{widetext}
Substituting Eqns. (\ref{eq:success_prob_arbi_meter_weak} - \ref{eq:int_f_partial_f_r_weak}) and Eq. (\ref{eq:int_f_f_deri}) into Eq. (\ref{eq:ftot_arbi_meter}), we have
\begin{widetext}
\be
F_{tot} = 4 \left\{\int dp |f'(p)|^2 -  \frac{1}{2} \left[\frac{(\cos\theta_i + \cos\theta_f)^2}{1+\cos\theta_i \cos\theta_f + \sin\theta_i \sin\theta_f \cos\phi_0} + \frac{(\cos\theta_i - \cos\theta_f)^2}{1-\cos\theta_i \cos\theta_f - \sin\theta_i \sin\theta_f \cos\phi_0}\right] \left| \int dp \tilde{f}(p) f'(p) \right|^2\right\}.
\label{eq:f_tot_arbi_meter_weak}
\ee
\end{widetext}
If $\int dp \tilde{f}(p) f'(p) = 0$, for example, $f(p)$ is a real function, $F_{tot}$ always equals $Q_j$. The Gaussian MA state discussed in the main text is a specific example of this situation. If  $\int dp \tilde{f}(p) f'(p) \neq 0$, $F_{tot}$ achieves its maximum value $Q_j$ when the pre- and post-selected QS states satisfy the condition
\be
\cos\theta_f \sin\theta_i - \cos \theta_i \sin \theta_f \cos\phi_0 = 0
\label{eq:max_cond_weak}
\ee

In the strong measurement limit with $g \to \infty$, we have $\int dp \tilde{f} (p+g) f(p-g) = 0$,  $\int dp \tilde{f} (p+g) f'(p-g) = 0$ and $\int dp \tilde{f}'(p+g) f'(p-g)=0$. Thus
\begin{widetext} 
\ben
p_d & = & \frac{1+\cos\theta_i \cos\theta_f}{2}, \label{eq:success_prob_arbi_meter_strong} \\
p_r & = & \frac{1-\cos\theta_i \cos\theta_f}{2}, \label{eq:fail_prob_arbi_meter_strong} \\
\int dp \left| \frac{\partial \vartheta_d (p,g)}{\partial g} \right|^2 & = & \frac{1+\cos\theta_i \cos\theta_f}{2} \int dp |f'(p)|^2, \label{eq:int_partial_d_strong} \\
\int dp \left| \frac{\partial \vartheta_r (p,g)}{\partial g} \right|^2 & = & \frac{1-\cos\theta_i \cos\theta_f}{2} \int dp |f'(p)|^2, \label{eq:int_partial_r_strong} \\
\int dp \tilde{\vartheta}_d (p,g) \frac{\partial \vartheta_d(p,g)}{\partial g} & = & \frac{\cos\theta_i + \cos\theta_f}{2} \int dp \tilde{f} (p) f'(p),
\label{eq:int_f_partial_f_d_strong} \\
\int dp \tilde{\vartheta}_r (p,g) \frac{\partial \vartheta_r(p,g)}{\partial g} & = & \frac{\cos\theta_i - \cos\theta_f }{2} \int dp \tilde{f} (p) f'(p),
\label{eq:int_f_partial_f_r_strong} 
\een
\end{widetext}
Sustituing Eqns. (\ref{eq:success_prob_arbi_meter_strong} - \ref{eq:int_f_partial_f_r_strong}) and Eq. (\ref{eq:int_f_f_deri}) into Eq. (\ref{eq:ftot_arbi_meter}), we have
\begin{widetext}
\be
F_{tot} = 4 \left\{\int dp |f'(p)|^2 -  \frac{1}{2} \left[\frac{(\cos\theta_i + \cos\theta_f)^2}{1+\cos\theta_i \cos\theta_f} + \frac{(\cos\theta_i - \cos\theta_f)^2}{1-\cos\theta_i \cos\theta_f}\right] \left| \int dp \tilde{f}(p) f'(p) \right|^2\right\}.
\label{eq:f_tot_arbi_meter_strong}
\ee
\end{widetext}
Again if $\int dp \tilde{f}(p) f'(p) = 0$, $F_{tot}$ always equals $Q_j$. If $\int dp \tilde{f}(p) f'(p) \neq 0$, $F_{tot}$ achieves its maximum value $Q_j$ when $\cos\theta_f =0$ with the post-selected QS state $|\psi_i\rangle = (|-1\rangle \pm e^{i\phi_i} |+1\rangle)/\sqrt{2}$. 

The above results show that the precisions one can achieve with weak and strong measurement  are the \textit{same} for an arbitary MA state when both measurements are optimized. 

If one retains only the information in the successfully post-selected meter state, $F_{tot} = p_d Q_d$. In the weak measurement limit $g \to 0$,
\begin{widetext}
\be
p_d Q_d = 4\left[ \frac{1+\cos\theta_i \cos\theta_f - \sin\theta_i \sin\theta_f \cos\phi_0}{2} \int dp |f'(p)|^2 - \frac{(\cos\theta_i + \cos\theta_f)^2 + \sin^2 \theta_i \sin^2 \theta_f \sin^2 \phi_0}{2(1+\cos\theta_i \cos\theta_f + \sin\theta_i \sin\theta_f \cos\phi_0)} \left| \int dp \tilde{f}(p) f'(p) \right|^2 \right].
\label{f_partial_arbi_meter_weak}
\ee
\end{widetext}
Similarly in the strong measurement limit $g \to \infty$, 
\begin{widetext}
\be
p_d Q_d = 4\left[ \frac{1+\cos\theta_i \cos\theta_f}{2} \int dp |f'(p)|^2 - \frac{(\cos\theta_i + \cos\theta_f)^2 }{2(1+\cos\theta_i \cos\theta_f)} \left| \int dp \tilde{f}(p) f'(p) \right|^2 \right].
\label{f_partial_arbi_meter_strong}
\ee
\end{widetext}
Both Eq. (\ref{f_partial_arbi_meter_weak}) and Eq. (\ref{f_partial_arbi_meter_strong}) are generally smaller than $Q_j$. Yet if $\int dp \tilde{f}(p) f'(p) = 0$, both of these equations can reach $Q_j$ with the conditions discussed in the main text.

\subsection{Quantum FI of the joint system-meter state in Eq.~(10)}

As shown in Sec.~\ref{sec:f_tot}, post-selection on the QS state followed by the measurement on the MA state can be considered as a measurement on the joint state $|\Psi_j\rangle$. Therefore, we can estimate the quantum FI of $|\Psi_j\rangle$, which will give us an upper bound on the precision, though it may not be achievable, in that $|\psi_f\rangle$ may not be the optimal.

We have
\ben
\frac{d|\Psi_j\rangle}{dg} & = & \cos \frac{\theta_i}{2}|-1\rangle (-i\alpha e^{-ig} \hat{a}^{\dag} )|\alpha e^{-ig}\rangle \nonumber \\
    && + \sin \frac{\theta_i}{2} e^{i\phi_i} |+1\rangle (i \alpha e^{ig} \hat{a}^{\dag})|\alpha e^{ig}\rangle \nonumber \\
& = & -i\alpha \hat{a}^{\dag} \left(e^{-ig} \cos \frac{\theta_i}{2}|-1\rangle |\alpha e^{-ig}\rangle \right.\nonumber \\
    && \left. - e^{ig} \sin \frac{\theta_i}{2} e^{i\phi_i} |+1\rangle |\alpha e^{ig}\rangle\right)
\een
then
\begin{equation}
\left(\frac{d\langle\Psi_j|}{dg}\right) \left(\frac{d|\Psi_j\rangle}{dg}\right) = n^2 + n.
\end{equation}
and
\begin{equation}
\langle \Psi_j |\left(\frac{d|\Psi_j\rangle}{dg}\right) = -i|\alpha|^2 \left(\cos^2 \frac{\theta_i}{2} - \sin^2 \frac{\theta_i}{2}\right)
= -i n \cos\theta_i.
\end{equation}
So we have the quantum FI of the joint state
\begin{eqnarray}
Q_{j} & = & 4\left[\left(\frac{d\langle\Psi_j|}{dg}\right) \left(\frac{d|\Psi_j\rangle}{dg}\right) - \left|\langle \Psi_j |\left(\frac{d|\Psi_j\rangle}{dg}\right)\right|^2 \right] \nonumber \\
& = & 4n^2 \sin^2 \theta_i + 4n.
\label{eq:qfi_joint}
\end{eqnarray}

\subsection{Quantum and classical FI for weak measurement in phase space}

The probability of successful post-selection $p_d$, that is, of obtaining the state $|\Phi_d\rangle$, is being given by
\begin{equation}
p_d = \frac{1+\mathcal{A}\cos(\mathcal{B}-2g)+\mathcal{C}}{2},
\label{eq:f}
\end{equation}
where $\mathcal{A} = \sin\theta_i \sin\theta_f \exp(-2n\sin^2 g),\mathcal{B}=n \sin 2g + 2g+\phi_0,$ $\mathcal{C}=\cos\theta_i\cos\theta_f,$ and $n=|\alpha|^2.$ The classical FI in this post-selection process is
\be
F_p = \frac{4 n^2 \mathcal{A}^2 \sin^2 \mathcal{B}}{1-\left[\mathcal{A}\cos(\mathcal{B}-2g)-\mathcal{C}\right]^2}.
\label{eq:info_post_phase}
\ee
The quantum FI for the state after successful post-selection is
\begin{widetext}
\be
Q_d  =  \frac{4}{p_d} \left\{ \frac{n}{2} (1+\mathcal{C} - \mathcal{A}\cos\mathcal{B})
    + \frac{n^2}{2} \left[1+\mathcal{C} - \mathcal{A}\cos(\mathcal{B}+2g)\right]
 - \frac{1}{p_d} \frac{n^2}{4} (\cos^2 \theta_i + \cos^2 \theta_f + 2\mathcal{C} + \mathcal{A}^2\sin^2\mathcal{B})\right\}.
\ee
The quantum FI for the state after failed post-selection is
\be
Q_r  =  \frac{4}{1-p_d} \left\{ \frac{n}{2} (1-\mathcal{C} - \mathcal{A}\cos\mathcal{B}) + \frac{n^2}{2} \left[1-\mathcal{C} + \mathcal{A}\cos(\mathcal{B}+2g)\right]
 - \frac{1}{1-p_d} \frac{n^2}{4} (\cos^2 \theta_i + \cos^2 \theta_f - 2\mathcal{C} + \mathcal{A}^2\sin^2\mathcal{B})\right\}.
\ee
\end{widetext}

\end{document}